\documentclass{article}
\usepackage{authblk}
\usepackage[utf8]{inputenc}
\usepackage[total={6in, 8in}]{geometry}

\usepackage{setspace}
\usepackage{dcolumn}
\usepackage{hyperref}
\usepackage{graphicx}
\usepackage{color}
\usepackage{multirow}

\title{Does putting your emotions into words make you feel better? Measuring the minute-scale dynamics of emotions from online data.}
\author[a,*]{Rui Fan}
\author[b]{Ali Varamesh} 
\author[c]{Onur Varol}
\author[b]{Alexander Barron}
\author[d]{Ingrid van de Leemput}
\author[d]{Marten Scheffer}
\author[b,d,e,*]{Johan Bollen}

\affil[a]{Beihang University, Beijing, P. R. China}
\affil[b]{Center for Complex Networks and Systems, Indiana University, Bloomington, IN}
\affil[c]{Center for Complex Network Research, Northeastern University, Boston, MA}
\affil[d]{Wageningen University, Wageningen, The Netherlands}
\affil[e]{Cognitive Science Program, Indiana University, Bloomington IN}
\affil[*]{Correspondence to: jbollen@indiana.edu \& buaafanrui@gmail.com}

\begin{document}

\maketitle

\begin{abstract}
Studies of affect labeling, i.e.~putting your feelings into words, indicate that it can attenuate positive and negative emotions. Here we track the evolution of individual emotions for tens of thousands of Twitter users by analyzing the emotional content of their tweets before and after they explicitly report having a strong emotion. Our results reveal how emotions and their expression evolve at the temporal resolution of one minute. While the expression of positive emotions is preceded by a short but steep increase in positive valence and followed by short decay to normal levels, negative emotions build up more slowly, followed by a sharp reversal to previous levels, matching earlier findings of the attenuating effects of affect labeling. We estimate that positive and negative emotions last approximately 1.25 and 1.5 hours from onset to evanescence. A separate analysis for male and female subjects is suggestive of possible gender-specific differences in emotional dynamics.
\end{abstract}

\section*{Introduction}

When you have a feeling, at what point do you become aware of it and tell your friends? Does expressing the feeling make you feel better or worse? How long does the feeling last? Studies of affect labeling, i.e.~putting your feelings into words, indicate that it can attenuate positive and negative emotions through a variety of autonomic, neural, social, or behavioral mechanisms ~\cite{torre2018}. When subjects are shown emotionally evocative images and asked to report their emotions, affect labeling reduced their distress or negative emotions \cite{lieberman:2011,constantinou:2014,taylor2003,thomassin2012}. Affect labeling has also been shown to lessen anxiety and fear towards phobias~\cite{kircanski2012,niles2015affect,niles2016writing}. Interestingly, affect labeling exerts its effects in spite of common expectations to the contrary\cite{lieberman:2011}. It may thus be considered an implicit emotion regulation mechanism in the sense that it exerts its effects without subjects deliberately applying it to regulate their emotions nor believing in its efficacy as an emotion regulation technique.

A number of empirical methods have been employed to measure the effects of affect labeling in the lab. Subjects can be queried about their emotional experience ~\cite{lieberman2007,kircanski2012} after the presentation of an emotion-evoking stimulus, using introspective measures, e.g.~survey instruments and diaries, where subjects (periodically) self-report their emotional state~\cite{mauss:measures2009}. Biophysical measurements of emotions~\cite{mauss:measures2009,kahneman2006,probst:emotion2016,phan2002functional,ochsner2002rethinking,fossati2003search,andreassi2013psychophysiology,nummenmaa2014bodily}, e.g.~measurements of facial musculature, in addition to brain scanning techniques can record individual reactions to emotionally evocative stimuli~\cite{mcrae2008gender,phan2002functional,ochsner2002rethinking,koelsch2006investigating,prasad:2018}. These measurements may involve extensive experimental and instrumental manipulation which may introduce measurement and observation challenges~\cite{mauss:measures2009}.

Here we measure the dynamics of \textit{naturally} occurring (i.e.~not experimentally induced) individual emotions and their spontaneous expression in online language~\cite{shariff2011} at the resolution of minutes for a very large sample of Twitter users. In particular, we find cases where users state ``I feel ...'' which we consider a case of affect labeling. We subsequently observe whether those expressions of a positive or negative valence emotion were associated with an intensification or attenuation of the emotion at different points in time \emph{before and after} the statement.

Our results align with, and extend, existing work in the area of affect labeling \cite{torre2018}. We observe rapid reductions of negative emotions and a less rapid reduction of positive emotions, immediately after their expression in language. The effect generalizes across most subjects indicating it is not the result of differences in personality or social dispositions. Since subjects were not aware of this research at the time they posted their tweets and were reporting ``naturally'' occurring emotions, our study may represent the first \textit{``in vivo''} observation of the effects of affect labeling on implicit emotional regulation.

\section*{Data and methods}

We obtained the individual Twitter timelines of 665,081 \emph{randomly selected} subjects, each timeline containing a longitudinal record of all tweets (up to a maximum of the 3,200 most recent) posted by the individual over time. Each tweet is marked by the date at which it was posted, the subject's user handle serving as a personal identifier, and the text content of the tweet (by design limited to a maximum of 140 characters). In total, our database contains 1,150,447,758 (approximately 1.15 Billion) tweets across all individual timelines.

\begin{figure*}[h!]
    \centering
    \includegraphics[width=0.9\textwidth]{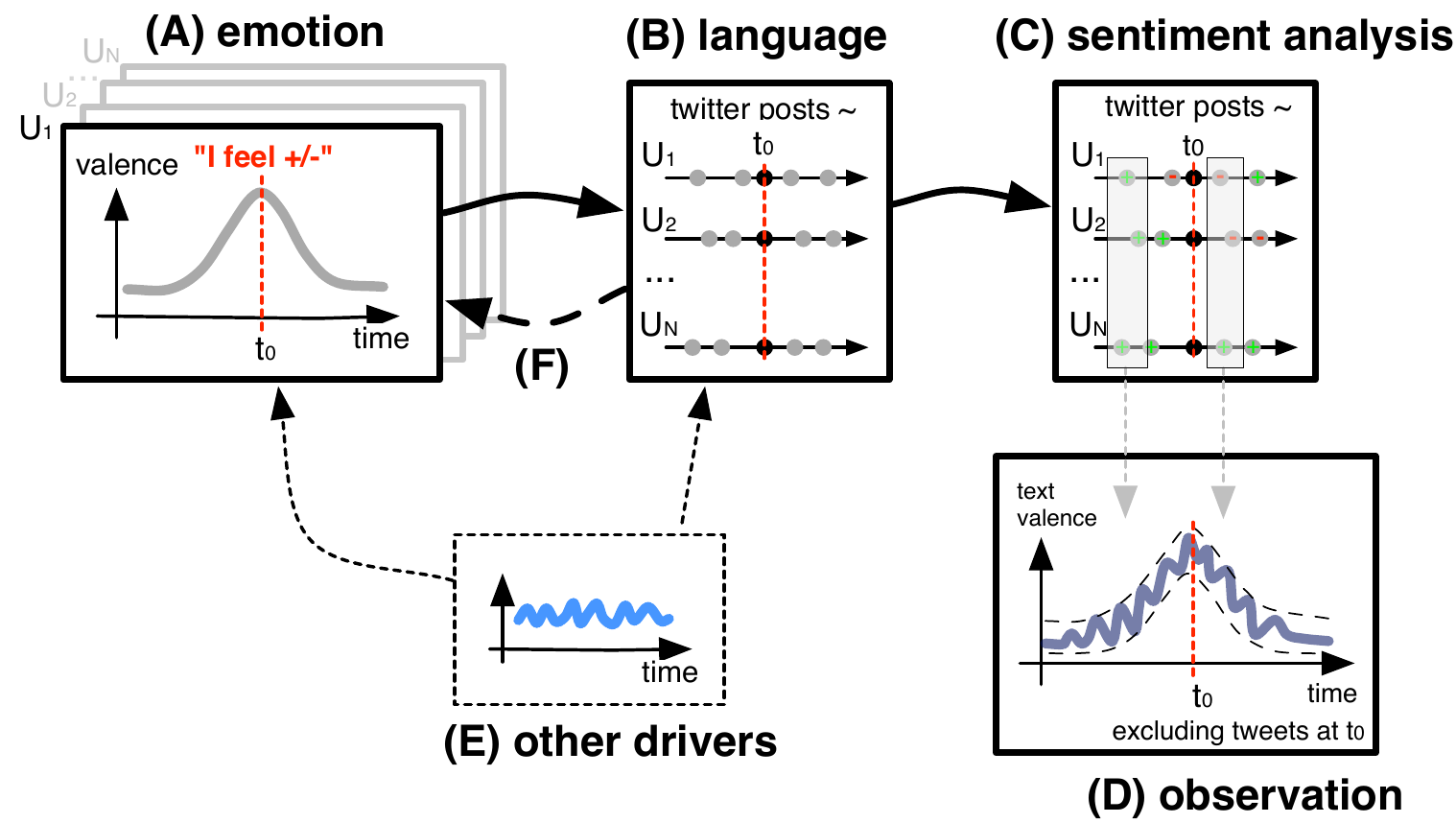}
    \caption{\label{user_valence}When subjects explicitly report having an individual emotion, can we observe its evolution over time from changes in their language? We collect the individual Twitter timelines of $N$ subjects that explicitly expressed the experience of having a positive or negative emotion at a specific time $t_0$ (A). We align the timelines according to their $t_0$ value for all subjects (B). If subjects are indeed experiencing an emotion (A), and it is of non-zero duration, and it biases the subject's language (B), then an off-the-shelf text sentiment analysis algorithm should detect changing text valence levels (C) at different points in time shortly before and after $t_0$. When we aggregate these valence levels across all subjects we map how emotions typically evolve over time relative to their expression (D). Emotions and language can be biased by other drivers (E), such as events, personal experience, and dispositions, but these effects will be randomized across subjects (A,B). Finally, since language can interact with emotions (F) we may also observe this interaction in the timing of valence changes in (D) relative to the time of the explicit emotional expression at $t_0$.}
\end{figure*}

As shown in Fig. \ref{user_valence} we then use a two-level approach. First, we identify a cohort of subjects that at some time explicitly and unambiguously reported having a positive or negative emotion through affect labeling. Second, we analyzed the text valence of the tweets that were submitted by these same subjects before and after the expression with an off-the-shelf sentiment analysis tool to trace the evolution of the reported emotion.

\subsection*{Detecting affect labeling in tweets}

Instances in our individual timelines where subjects explicitly and unambiguously report their emotional state are identified by their stating ``I feel ...'', and two grammatical variations: ``I am feeling'', or ``I'm feeling'', an approach developed earlier to detect personal emotions through self-reports \cite{Bollen:twitter2011,Bollen:modeling2011,yang:life2016}. We use the adverb or adjective that follows ``I feel'' as an indication of the statement's emotional valence direction. For example, ``I feel so \emph{unhappy}'' is taken to mean that the subject is having a strong negative emotion since ``unhappy'' is deemed to have negative valence. Here, we focus on valence since it is a dominant component in most descriptive and nominal models of human emotions~\cite{darwin1872,russel1980,russel1977,mehrabian1980,plutchik1997,ekman1999,lang2008}.

Note that we use the ``I feel'' expression, i.e. the affect labeling statement, only as a binary diagnostic, i.e.~to determine whether or not the subject reported a positive or negative emotion at that time. The statement is not used to analyze the emotion itself nor to capture the fullest possible spectrum of human emotions.

Consequently, we look only for the clearest and most unambiguous indications that a positive or negative emotion did indeed take place. Although many adjectives or adverbs may be used in ``I feel '' expressions to indicate high or low valence emotions, they can also express specific (combinations of) different emotions or physical states. For example, ``I feel sick'' is indicative of a negative valence state, but it could also mean that one is emotionally ``disgusted'' or physically unwell. Including such adjectives would reduce the reliability and validity of our results since they would reflect an unknown combination of emotions and valence levels. It is therefore methodologically sound practice to choose the most limited set of adjectives that also most unambiguously indicate that an emotion actually took place.

Hence, we selected only 10 of the most frequent and explicit valence-related adjectives or adverbs in all ``I feel'' statements in our data, specifically, for positive valence: ``good'', ``happy'', ``great'', ``awesome'' and for negative valence: ``bad'', ``unhappy'', `sad'', ``terrible'', ``horrible'', ``awful'', along with a few boosters (``so'', ``very'', etc). The adjective ``unhappy'' was included as a counterpoint to the positive ``happy''.

This selection of 10 adjectives represents more than 98\% of all "I feel " statements in our data that unambiguously pertain to high or low valence only (excluding those that do not, such as ``I feel weird'' or ``I feel sick''). Adding more adjectives and adverbs could thus only slightly increase our sample, but at the expense of reducing the validity of our observations (see Supplemental Information).

In total we find 109,943 expressions containing the listed adjectives. As mentioned, we use the valence of the adjective or adverb of the expression to separate our timelines into a positive group (N=42,627) and a negative group (N=67,316). The tweet volume per time period in Fig.~\ref{window_volume} similarly shows a greater number of negative statements.

To avoid traces of explicit or intentional emotional statements biasing our results, we also exclude: (1) the self-report tweet itself and (2) any timelines that contain more than 1 self-report statement within 48 hours of each other (see Supplemental Information for a full overview).

\subsection*{Sentiment analysis of tweets posted before and after the affect labeling}

For all subjects in both groups we retrieve the tweets posted in their individual timelines 6 hours before to 6 hours after the time of the emotional expression, that is a total of 12 hours surrounding the expression. We center all resulting 12 hour timelines on the time of the expression, referred to as $t_0$.

We then use an accurate, off-the-shelf Twitter sentiment analysis algorithm, i.e.~VADER~\cite{hutto2014vader,Ribeiro2016},  to assign each individual tweet (again excluding the emotional statement itself) a numerical valence score between -1 (very negative) and +1 (very positive). VADER recognizes 7,516 terms of the most frequently used English words, that were each rated by 10 independent human raters to assess their Valence value. VADER recognizes negations, hedging, boosters, colloquial language, style, jargon, and abbreviations that are commonly used on Twitter. It responds only to English, hence our analysis automatically rules out non-English users and content.

Tweets are individually posted by our subjects at irregular points in time within the mentioned 12 hour period that is centered on the emotional expression at time $t_0$. To observe changes in emotional valence relative to $t_0$ we group tweets in discrete 1 minute time windows according to the time $t_0 \pm k$ they were posted, from 6 hours before $t_0$ to 6 hours after $t_0$, i.e. $k \in [-360^\prime,+360^\prime]$ minutes, and we do so separately for both positive and negative groups.

Each 1 minute window contains a set of tweets that were posted at that given distance $k$ from $t_0$, producing a distribution of VADER valence values. To be able to draw inferences about significant changes in Valence levels over time, we obtain converging evidence through two distinct methods:

\begin{enumerate}
    \item We produce a time series of mean Valence values for each 1 minute window and apply a CUSUM~\cite{cusum} procedure that examines cumulative changes in the time series' variance to detect significant change points
    \item We bootstrap mean valence values and compare the resulting distribution to that produced by a null-model of a random sample of tweets with similar weekly, circadian, and diurnal features as the tweets in the time window under consideration~\cite{Golder1878}.
\end{enumerate}

We thus obtain two sources of evidence with respect to significant changes in Valence levels.\\

Finally, since men and women may differ in their emotional responses and expressions~\cite{kring1998sex,mcduff2017}, we separate our timelines into male and female sets using a gender classifier (accuracy=86.4\% and AUC=0.916) that was trained using 1,744 manually labeled social media accounts. This results in two positive and negative timelines with all subjects combined, as well as 4 timelines differentiated by gender (female and male) and emotion (negative and positive).

\section*{Results}

\begin{figure*}[h!]
\begin{center}
    \includegraphics[width=6in]{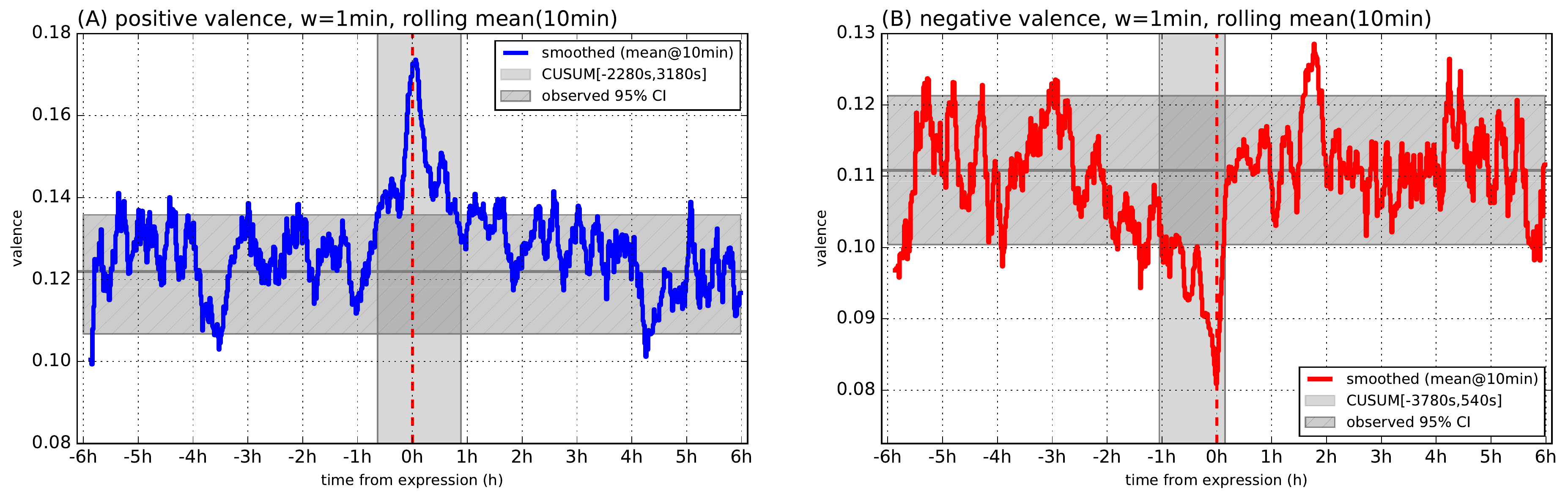}
    \caption{\label{timeseries_CUSUM10} 
Time series of observed valence values across all subjects (positive N=42,627 and negative N=67,316), 6 hours before and after subjects make an explicit expression of experiencing an emotion at time $t_0$ show statistically significant changes in subjects' Valence levels. Mean Valence values were computed for all tweets posted in the same 1 minute intervals, smoothed by a rolling average in a 10 minute window. We use the first 3 hours to estimate a baseline 95\% CI for the entire time series. A CUSUM test is used to detect statistically significant change points in the resulting time series. A gray vertical bar marks where a CUSUM analysis indicates significant change point based on the time series' cumulative variance. The time series reveal a distinct pattern of change before the emotional affirmation: a positive ramp up and negative ramp down before the subjects' explicit positive and negative affirmations respectively, followed by a ramp down and ramp up of the subjects' emotional state afterwards. Positive and negative emotions seem to follow different trajectories; negative emotions have a longer ramp-up period, possibly starting at $t_0-2$hrs, but a faster recovery immediately after the emotional expression. This may point to the presence of an affect labeling effect in the expression of an emotion is followed by a reduction of negative valence (see Fig. \ref{fig:fitting}).}
\end{center}
\end{figure*}

As shown in Fig.~\ref{timeseries_CUSUM10}, we plot mean Valence values in 1 minute increments from 6 hours before and 6 hours after an emotional expression a time $t_0$. We smooth these time series by a rolling average of 10 minutes. Note that the emotional expressions themselves at time $t_0$ are \emph{excluded} from this analysis and that we do not analyze timelines that contain more than 1 explicit emotional expression within a 48 hour span.

The positive emotion group in Fig.~\ref{timeseries_CUSUM10}(A) shows a sharp peak of Valence values before and after $t_0$. This indicates that the subjects' language does reflect positive valence changes that match the subjects' positive explicit statement. Ramp-up and ramp-down of positive Valence levels seems to be rapid and symmetric around $t_0$, with a sharp peak located exactly at $t_0$, indicating that the peak of emotional levels coincides with the emotional expression, and decays immediately after. All other fluctuations in the time series fall within a 95\% CI established from the distributions of Valence values in the first 3 hours of the time series. Similar CIs are observed when we sample the last 3 hours of the time series instead.

To confirm that the observed changes are significant we conduct a CUSUM test, which measures cumulative changes in the time series' variance. The results indicate statistically significant changes in Valence values from $t_0 - 38m$ to $t_0 + 53m$ (grey band in Fig. \ref{timeseries_CUSUM10}(A)), indicating that positive emotions typically last 92 minutes or one hour and 32 minutes.

The negative emotion group in Fig.~\ref{timeseries_CUSUM10}(B) exhibits an equally sharp change in Valence levels surrounding the emotional expression at time $t_0$, but with a slower and longer ramp-down before a sharp negative peak at $t_0$, and a fast reversal to the long-term median within 10 minutes after the emotional expression. 

A CUSUM test again confirms this observation for the negative emotion revealing a period of statistically significant changes in negative valence from $t_0 - 63m$ to $t_0 + 9m$, indicating that negative emotions start earlier than positive emotions, take longer to crest, and end soon after the emotional expression, with a typical duration of 73 minutes or about 1 hour and 13 minutes.

We furthermore note interesting differences between positive and negative baseline levels of valence. The baseline valence of subjects with positive emotions runs in a slightly higher range, 95\% CI $[0.107, 0.136]$, than that of subjects with negative emotions, 95\% CI $[0.100,0.121]$. This may be indicative of structural differences in our sample of positive and negative timelines, i.e.~negative emotions and their expression may be more prevalent among negatively inclined individuals whereas positive emotions and their expression may be more prevalent among more positively inclined individuals. The construction of our null-model below will specifically take into account such potential bias.

\subsection*{Modeling emotional dynamics}

\begin{figure*}[h!]
    \centering
    \includegraphics[width=\textwidth]{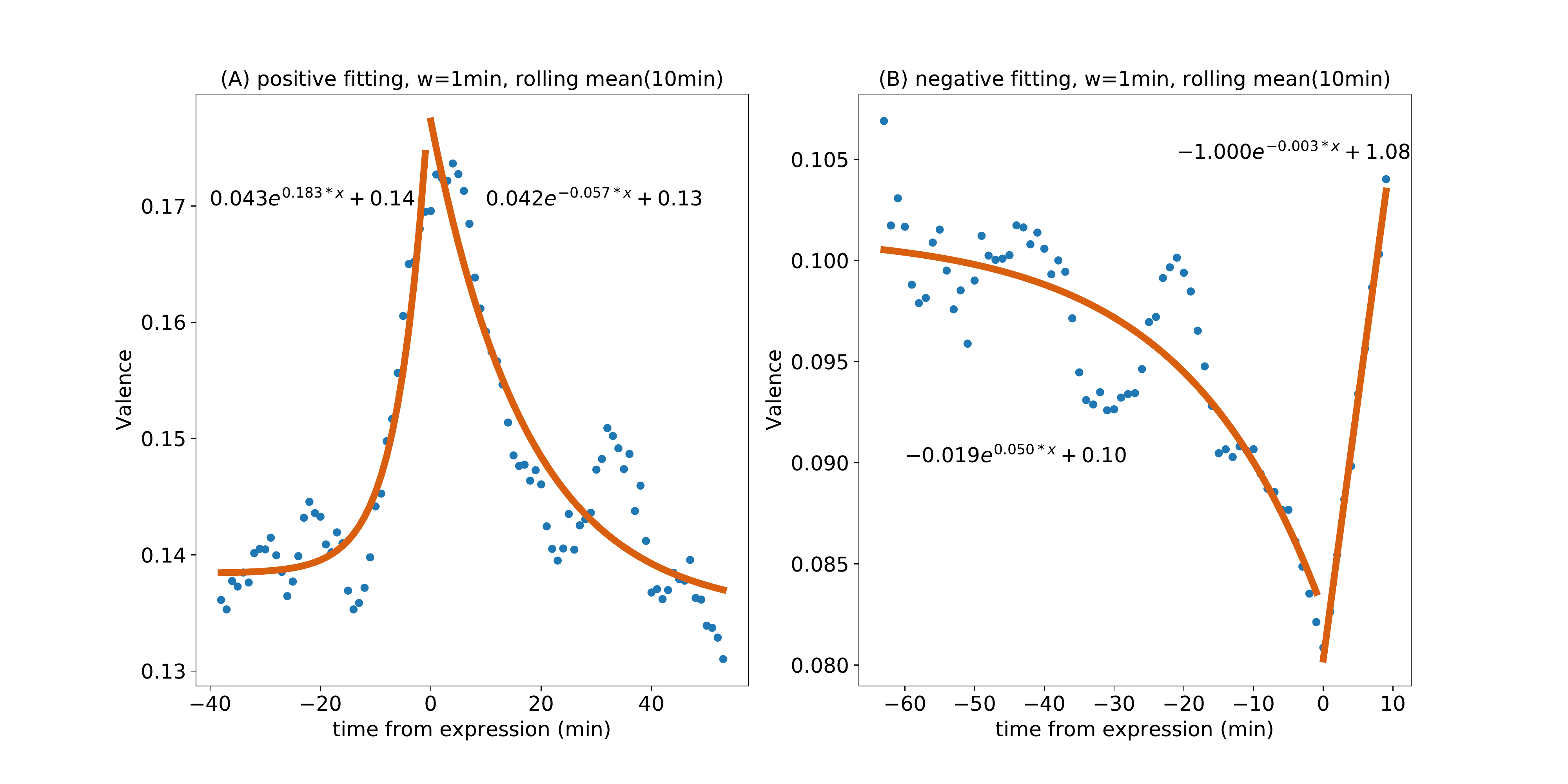}
    \caption{Curve fitting results of the smoothed mean Valence values. The ramp-up and ramp-down of both time series are best fitted by 2 separate exponential growth and decay functions (Least Squares Method) before and after the emotion affirmation point respectively (see Table \ref{fit_errors}). Both positive and negative emotions exhibit rapid and accelerating waxing and waning.}
    \label{fig:fitting}
\end{figure*}

To quantitatively confirm our observations above and model the dynamics of positive and negative emotions, we apply a least squares method to fit the ramp-up and ramp-down of the positive and negative time series. As can be seen in Fig.~\ref{fig:fitting}, the positive ramp-up and ramp-down periods are best fitted with a separate (1) exponential growth function $f(t \in [-38m, -1m])=0.043e^{0.183t}+0.14$ and (2) an exponential decay function $f(t \in [0m, 53m])=0.042e^{-0.057t}+0.13$ respectively. For the negative emotion, the growth before $t_0$ and the decay period after $t_0$ is also best fitted with a separate exponential growth function $f(t \in [-63m, -1m])=-0.019e^{0.050t}+0.10$ and exponential decay function $f(t \in [0m, 9m])=-e^{-0.003t}+1.08$ respectively.

\begin{table}[]
    \centering
    \begin{tabular}{ l | l | l }
    \textbf{Positive:} fitted functions    &  SSE    &  Parameters              \\\hline
    $f(t)=Ae^{\lambda t}+b$ (2 Exponentials)    &   \textbf{0.001389*}  & \multicolumn{1}{l}{\begin{tabular}[c]{@{}l@{}}1) $A=0.043$, $\lambda=0.183$, $b=0.14$\\ 2) $A=0.042$, $\lambda=-0.057$, $b=0.13$\\\hline\end{tabular}} \\
    $f(t)=\frac{A}{\pi}[\frac{\sigma}{(t-\mu)^2+\sigma^2}]$ (Lorentzian Model)     &   0.005509     & $\sigma=100.38$, $\mu=6.44$, $A=49.32$      \\
    $f(t)=\frac{A}{\sigma \sqrt{2\pi}}\exp{-\frac{(t-\mu)^2}{2\sigma^2}}$ (Gaussian)           &   0.005642 & $\sigma=74.79$, $\mu=6.50$, $A=29.23$    \\
    $f(t)=at^2+bx+c$ (Quadratic Model)      &   0.005770  & $a=-1.26e^{-5}$, $b=0.00016$, $c=0.15$   \\\hline
    \multicolumn{3}{c}{}\\
    \multicolumn{3}{l}{\textbf{Negative:} fitted functions}\\\hline
    $f(t)=Ae^{\lambda t}+b$ (2 Exponentials)    &   \textbf{0.000414*}  & \multicolumn{1}{l}{\begin{tabular}[c]{@{}l@{}}1) $A=-0.019$, $\lambda=0.050$, $b=0.10$\\ 2) $A=-1.000$, $\lambda=-0.003$, $b=1.08$\\\hline\end{tabular}}  \\
    $f(t)=at^2+bx+c$ (Quadratic Model)      &   0.001298 & $a=1.55e^{-6}$, $b=-0.00010$, $c=0.091$    \\
    $f(t)=\frac{A}{\pi}[\frac{\sigma}{(t-\mu)^2+\sigma^2}]$ (Lorentzian Model)     &   0.001317 & $\sigma=106.81$, $\mu=-1025.39$, $A=2821.67$    \\
    $f(t)=\frac{A}{\sigma \sqrt{2\pi}}\exp{-\frac{(t-\mu)^2}{2\sigma^2}}$ (Gaussian)           &   0.001320  & $\sigma=2034.21$, $\mu=-8213.89$, $A=1.60$  \\\hline
    \end{tabular}
    \caption{Sum of Squared Errors (SSE) for multiple functions that were fit againt the ramp-up and ramp-down of positive and negative Valence, in rank-order of lowest SSE (marked *). Best fit is achieved with two separate exponential growth and decay functions, i.e.~ one before and one after $t_0$, as shown in Fig. \ref{fig:fitting}.}
    \label{fit_errors}
\end{table}

Based on the fitted curves, we calculate the half-life of both emotions. Given the peak value of an emotion $p$ and the emotion value at the end of emotional period $e$ (average emotion score), we define the half-life as $t_{(p+e)/2}-t_p$, indicating the time it takes for the emotion to decay from the peak to half of its baseline value. By this definition, the half-life of positive and negative emotion is 11 and 5 minutes, respectively, indicating a swift return from peak levels to the Valence baseline.

It is suggestive that most ramp-ups and declines of the subjects' emotional states, positive as well as negative, are best fit with exponential trajectories when compared to other functions (see Table \ref{fit_errors}). This may be indicative of the presence of feedback loops in the subjects' emotional system that were affected or interrupted by the subjects' explicit expression at $t_0$.

\subsection*{Mean Valence Confidence Intervals vs. a Null-model}

\begin{figure*}[h!]
\begin{center}
    \includegraphics[width=6in]{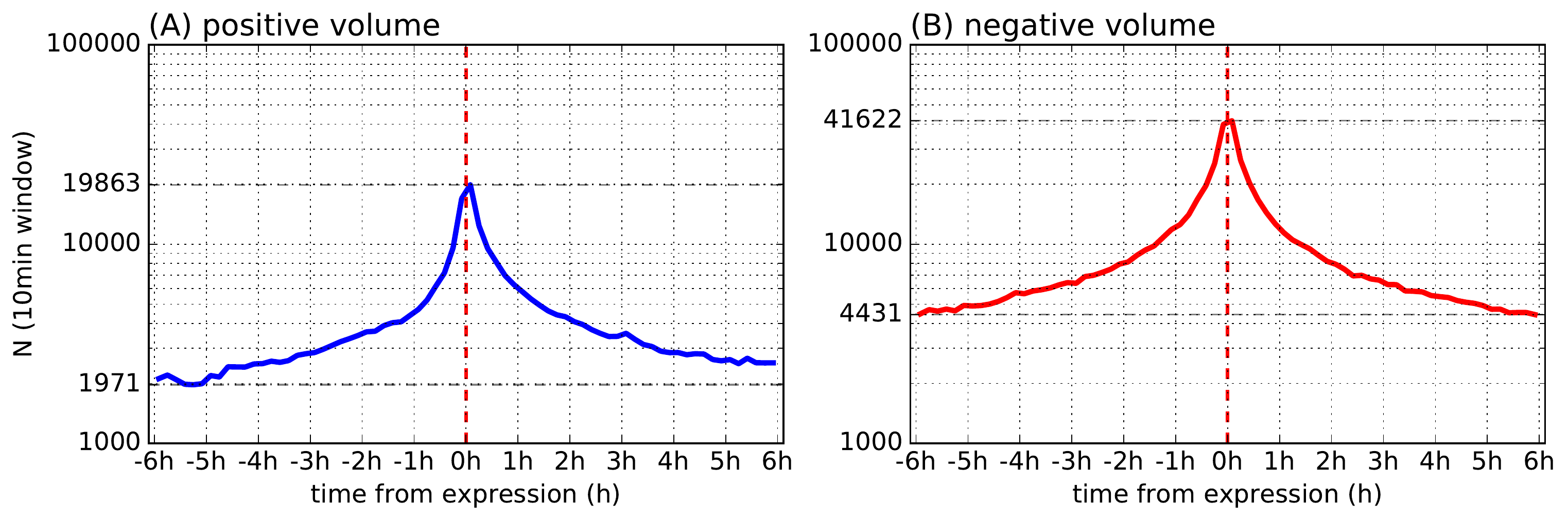}
    \caption{\label{window_volume} Number of tweets in analysis window over time for positive and negative groups. The sample of tweets in the positive group ranges from a minimum of 1,971 to a maximum of 19,863 tweets. The sample of tweets in the negative group ranges from a minimum of 4,431 to a maximum of 41,622 tweets. Note that similar to our timeline sample, we find a significant larger volume of tweets in the negative group vs.~the positive group.}
\end{center}
\end{figure*}

Each time window $w$ in our data contains a (changing) sample of tweets that were posted within a distance $k$ from $t_0$, i.e. the tweets posted in the time interval $[t_0-k-w, t_0-k]$. Since each tweet in that window has been assigned its individual Valence rating, mean Valence values for a window (such as shown in Fig.~\ref{timeseries_CUSUM10}) are thus calculated for a \emph{distribution} of $n$ tweet Valence values. The properties of this distribution can change in terms of sample size, location, and variance between subsequent time windows, complicating inferences about differences in Valence levels across the time series.  In addition, as shown in Fig. \ref{window_volume}, the volume of tweets for each time window can vary depending on the distance from $t_0$. The volume of tweets per time window is highest in the vicinity of $t_0$ since this is the time we center our timelines on.

Our estimates of mean valence at time $t$ may therefore be more or less uncertain depending on the underlying sample of tweets. To estimate the error of our estimation of mean valence at window $t$, we bootstrap 10,000-fold: (1) mean Valence values and (2) null-model mean Valence values. To increase our sample size, we group tweets in 10 minute windows. 

\begin{figure*}[ht!]
\begin{center}
    \includegraphics[width=6in]{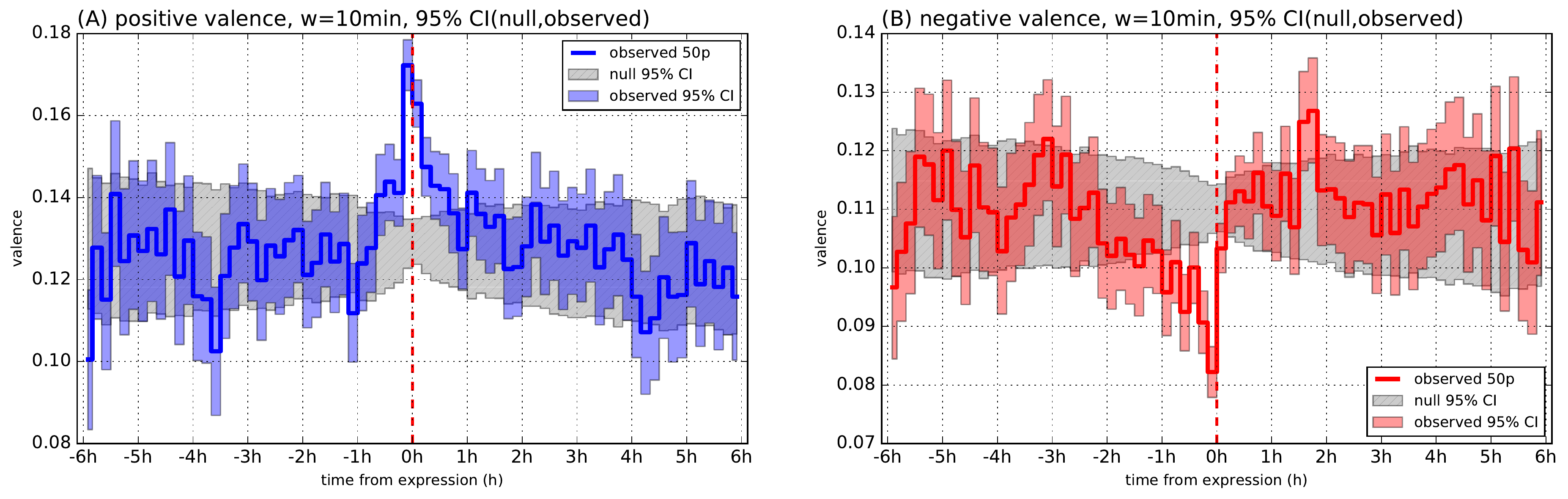}
    \includegraphics[width=6in]{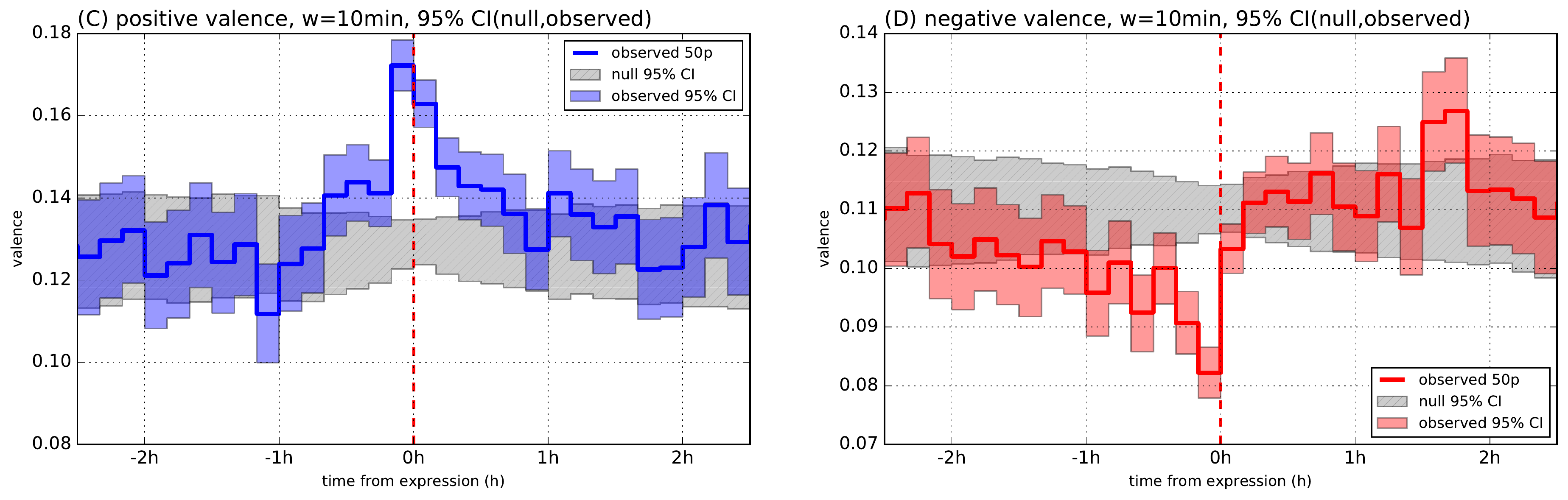}
    \caption{\label{timeseries_KS10} 
Time series of observed valence values at 10 minute intervals for positive and negative group show statistically significant changes in subjects' Valence levels before and after making an explicit expression of experiencing an emotion. Blue and red bands show 95\% CI of bootstrapped mean Valence levels. Gray bands show the 95\% CI of Valence values for a null-model of randomly chosen tweets (see Methods). Panels C and D show time series for the range $[-180,+180]$ minutes to show the details of the evolution of positive and negative valence surrounding $t_0$.
}
\end{center}
\end{figure*}

In Fig. ~\ref{timeseries_KS10} we chart the resulting median valence (50th percentile) and its 95\% Confidence Interval for the bootstrap replicate distribution in each time window alongside the null-model for each time window $w$. This null-model (see Methods) randomly samples tweets at the same distance $k$ from $t_0$ as $w$, with the same distribution of diurnal and circadian features (time of day and week day) as the observed tweets in $w$, from the positive and negative timelines respectively. As such it accounts for possible differences in subject dispositions between the positive and negative group, as well as possible circadian and diurnal biases.

The results of this comparison are shown in Fig.~\ref{timeseries_KS10}. The time series are displayed for discrete and adjacent windows of 10 minutes. Comparable results were obtained with 1, 5, and 15 minute windows. The 95\% CI of the estimated mean Valence levels are shown (red and blue bands) as well as the 95\% CI for the null-model estimates (gray band) for each window.

The CIs of observed valence overlap with those of the null-mode for most of the 12 hour period under consideration, with the exception of the period from -10m to +20m where we observe that the CIs of the observed mean positive Valence and those of the null-model do not overlap. This period overlaps with the CUSUM change point detection, however it is shorter, possibly due to applying the more strict criterion of non-overlapping 95\% CIs. We can draw a similar conclusion for the negative time series, i.e.~the 95\% CIs do not overlap from -40m to 0m, confirming a negative emotional period similar to that given by the CUSUM test. The span of the emotional period estimated by this strict criterion might be an underestimation, since areas where the CIs do overlap might still represent cases where Valence levels are significantly different from null-model levels.

\subsection*{Estimating emotion duration}

As shown in Table~\ref{emotion_period_estimation}, the CUSUM and 95\% CIs indicate different, yet overlapping emotional periods surrounding time $t_0$. The timing of the emotional period as indicated by CUSUM test indicates time ranges of elevated or depressed valence distributions from $k \in [-38,+53]$ for the positive emotion and $k \in [-63,+9]$ for the negative emotion. The CIs indicate a period of significant divergence in valence vs.~a null-model in the ranges of $k \in [-10, +20]$ and $k \in [-40,0]$ minutes for the positive and negative emotions respectively. We add a less strict criterion namely the time period when the time series remains continuously above or below the median surrounding $t_0$. This criterion indicate that the positive emotional period may range from $k \in [-48, 109]$ and the negative emotion period ranges from $k \in [-124,14]$.

\begin{table}[]
    \centering
    \begin{tabular}{l| rr | rr }
      Method                    &   $+$ Length  &   $+$Span  &   $-$ Length    &   $-$Span     \\\hline
          CUSUM                     &   $92^\prime$ & [-38,+53]       &   $73^\prime$  &  [-63,+9]             \\
      95\% CI                   &   $31^\prime$ &    [-10,+20]       &   $41^\prime$   & [-40,0]             \\
      $V_{[t_1,t_2]}<,>Q_2$     &   $158^\prime$    & [-48, 109]     &   $139^\prime$  &    [-124, 14]         \\\hline
      Average                   &   $94^\prime$ &    [-32,+61]       &   $85^\prime$   &    [-76,+8]            \\
    \end{tabular}
    \caption{Duration and span of changes in 
        Valence levels for positive and negative groups according to CUSUM, 95\% CI non-overlap, and
        length of continuous time period between which valence levels deviate from median ($Q_2$) before and after $t_0$ (all rounded to whole minutes).}
    \label{emotion_period_estimation}
\end{table}

Averaging the results of these three tests, we estimate that positive emotions range from $[-32,+61]$, i.e. a total duration of 94 minutes or 1.6 hours whereas negative emotions range from $[-76,+8]$, i.e. a total duration of 85 minutes or 1.4 hours. This indicates that negative emotions do not last longer than positive emotions, but they start earlier, have a significantly longer run-up, yet seem to decline and end more rapidly after their expression.

\subsection*{Robustness across subject samples}

We gauge the robustness of our results by determining the distribution and magnitude of \emph{peak} positive and negative valence levels for all individual subjects. We convert all individual subject time series $T$ to z-scores relative to its overall mean $\bar{x}(T)$ and standard deviation $\sigma(T)$,~i.e.~, the z-score of every valence value at time $k$ for time series $T$ is given by $Z(v_k) = \frac{v_k-\bar{x}(T)}{\sigma(T)}$. We then extract the largest and smallest z-scores in the emotional
periods detected by the above mentioned CUSUM procedure surrounding $t_0$ (see Table \ref{emotion_period_estimation}), to measure the magnitude of the positive and negative peak values across all subjects.

\begin{figure*}[h!]
    \centering
    \includegraphics[width=\textwidth]{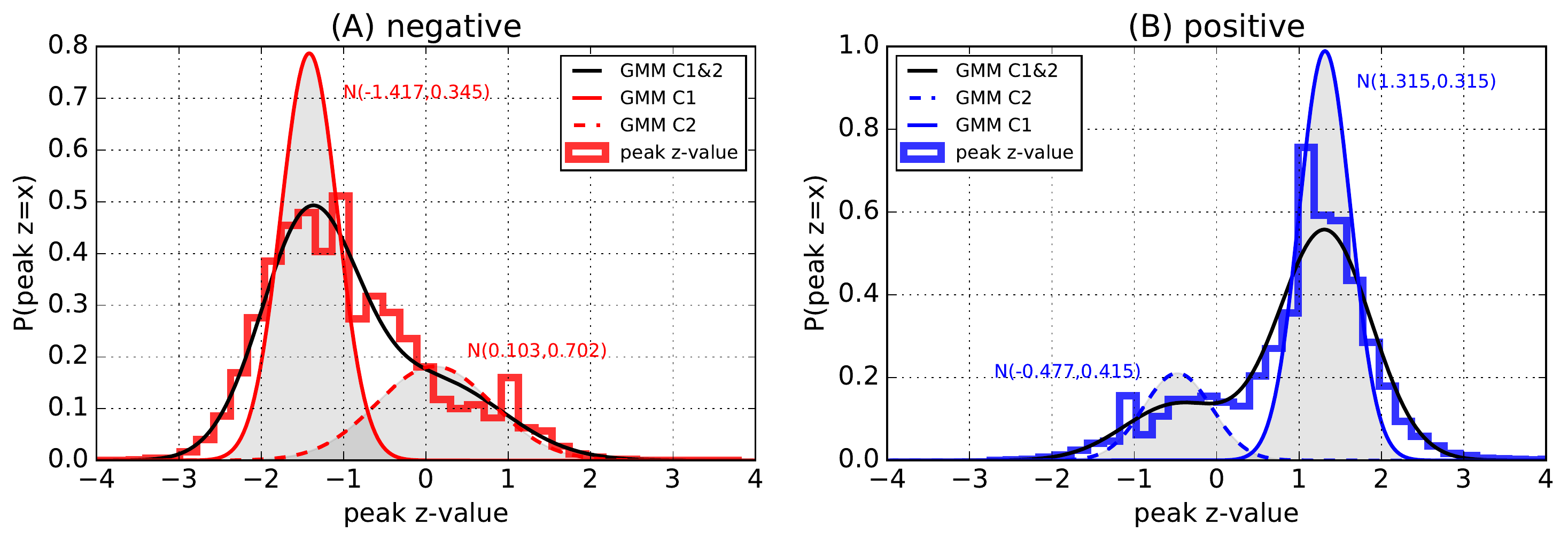}
    \caption{The distributions of peak z-scores of the valence time series for negative (A) and positive (B) emotions modeled by a 2-component Gaussian mixture model reveals that the majority of users exhibit a similar and significant emotional response matching the Valence of their self-report. Both emotions show a second Gaussian Mixture Model component that may correspond to a small sub-sample where the sentiment analysis tool (VADER) returns zero, insignificant, or contrary sentiment scores.}
    \label{fig:user_sampling}
\end{figure*}

The resulting distributions of peak z-values for the negative and positive emotions are shown in Fig. \ref{fig:user_sampling}(A,B). As expected, the distribution of the negative peak responses has a 95\% CI [-2.434, 1.285] and median of -1.073 (negative response), whereas the distribution of positive peak responses has a 95 \%CI of [-1.240, 2.417] and a median of +1.096 (positive response), indicating a large majority of negative and positive peak values are consistent with the Valence of the subjects' self-reports.

We then model each distribution with a 2-component Gaussian Mixture Model (GMM) since that number of components offers the most significant yield in Akaike information criterion (AIC) and  Bayesian information criterion (BIC) reduction. The estimated models provide weights and corresponding Gaussian components to describe underlying distribution as $P(x)=w_1\mathcal{N}_1+w_2\mathcal{N}_2$. Each Gaussian component is specified by a mean and standard deviation $\mathcal{N}(\mu, \sigma)$.  The negative emotion yields two components with weights $w_1=0.680$ and $w_2=0.320$ with components $\mathcal{N}_1(-1.417, 0.345)$ and $\mathcal{N}_2(0.103, 0.702)$, respectively. For the positive emotion 2 components were estimated, namely component C1 with weight $w_1=0.781$ and Gaussian parameters $\mathcal{N}_1(1.315, 0.315)$ and a second component C2 with weight $w_2=0.219$ and Gaussian parameters $\mathcal{N}_2(-0.477, 0.415)$. 

These results confirm that a strong and consistent emotional response generalizes across most subjects for negative and positive emotions, but both distributions do exhibit a positive and negative skew respectively. Although the first component C1 of the distribution of negative peak values is centered on negative values as expected, the second component C2 is focused on neutral peak values, indicating attenuated or neutral peak valence z-scores. The distribution of positive peak z-scores exhibits a symmetric pattern: the distribution is centered on positive peak values with the first component C1, but exhibits a negative skew, due to its second component (C2) that is centered around lower, more neutral peak sentiment values.

In most cases subjects do exhibit a strong positive or negative emotional response consistent with the Valence of the self-reported emotion, but a minority of cases has low, absent, or even inverted emotional peaks, possibly due to subject language preferences or the VADER sentiment analysis tool producing a neutral sentiment rating or failing to interpret the proper valence of irony, sarcasm, negation, or hedging. For negative and positive emotions we find that respectively 16.80\% and 17.89\% of peak values run contrary to the polarity of self-reports (positive peak value for negative emotion and vice versa). These results suggest that our observations of emotional responses around $t_0$ are most likely an \emph{underestimation} of the actual effect, since our data may include a sub-set of cases where the sentiment analysis yielded a neutral, attenuated, or inverted sentiment signal.

\subsection*{Male and female responses}

\begin{figure*}[h!]
\begin{center}
    \includegraphics[width=6in]{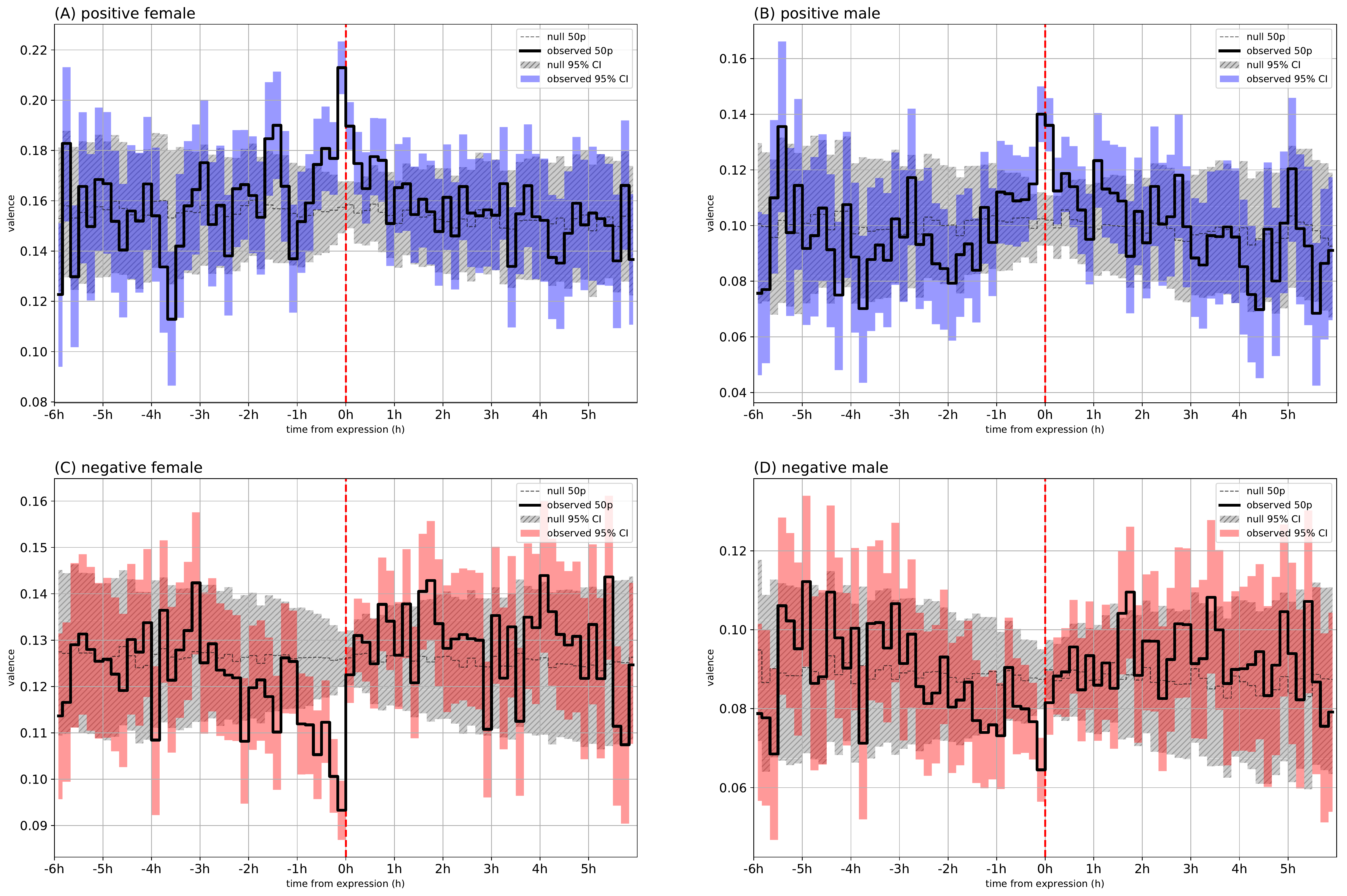}
    \caption{\label{timeseries_simple_femmal_10min} Gender differentiated time series of mean valence values in adjacent 10 minute windows before and after an explicit expression of having a positive or negative emotion at time $t_0$. Blue and red bands show 95\% CI of bootstrapped mean valence levels. Gray bands show the 95\% CI of valence for a null-model.}
\end{center}
\end{figure*}

Although the above results suggest a robust and generalizable emotional response across most subjects, when we separate our analysis for male and female subjects using a gender classifier, we do observe some differences in pattern, as shown in Fig.~\ref{timeseries_simple_femmal_10min}. Women seem to exhibit a higher positive and negative baseline than men do, with men exhibiting more variance of positive Valence levels than women, a more prolonged and more pronounced positive response, and possibly a reduced negative response with a slower recovery to baseline levels than women.

The effect in which negative emotions dissipate rapidly after the subjects' explicit expression ~\cite{pennebaker1995,moore2001} appears more pronounced for female subjects than for male subjects (on the basis of a visual inspection of the relevant time series), indicating that women may experience a stronger positive effect of expressing their emotions than men do or use different regulation mechanisms~\cite{ford:psych2017}.

We caution that observations with respect to baseline Valence levels may in part be shaped by implicit gender bias in sentiment analysis tools and lexicons~\cite{warriner2013norms}. However, this bias would not affect the pattern of relative changes that we observe in our time series.

\begin{table*}
\centering
\medskip
\begin{tabular}{l|r|r}
\hline
                & 95\% CI (10min)                   & CUSUM (1min)              \\
Gender, emotion & duration:[span]                   & duration:[span]           \\\hline
Female+         & 20$^\prime$: [-10, +10]           & 94$^\prime$:[-41, +53]    \\
Male+           & 20$^\prime$:[-10, +10]            & 97$^\prime$:[-48, +49]    \\
Female-         & 40$^\prime$:[-40, 0]              & 48$^\prime$:[-41, +7]     \\
Male-           & 10$^\prime$:[-10,0]               & 40$^\prime$:[-32, +8]     \\\hline
\end{tabular}
\caption{\label{tab:detection}The results of CUSUM and 95\% CI divergence (non-overlap) analysis of beginning and end of positive and negative emotional period, separated by gender. Start and end points are expressed in whole minutes. CI-based estimates are calculated at 10 minute intervals.}
\end{table*}

When we apply the same CUSUM and examine the timing of non-overlapping 95\% CIs to detect significant changes in Valence levels in the separate male and female time series, we do find overlapping periods of diverging Valence around $t_0$ in all time series (see Table \ref{tab:detection}). This indicates that we detect a significant emotional signal before and after $t_0$ for both male and female subjects separately. However, this observation pertains to emotion magnitude and duration, not whether the pattern of the emotional change evolves differently for male and female subjects.

\begin{figure*}[h!]
\begin{center}
    \includegraphics[width=6in]{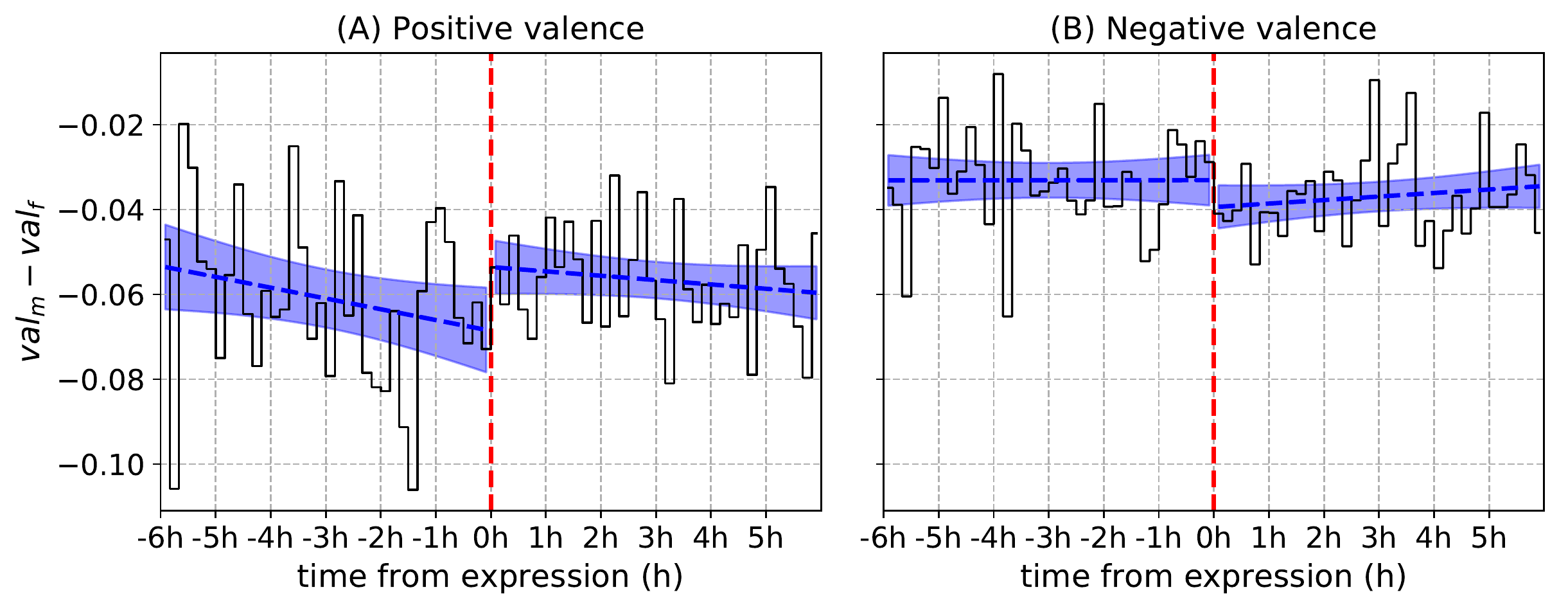}
    \caption{\label{gender_difference} Differences between male ($val_m$) and female ($val_f$) time series of mean valence values at 10 minute intervals for positive (A) and negative (B) valence. We applied regression discontinuity analysis to the $val_m - val_f$ time series before and after the explicit expression of experiencing an emotion at time $t_0$.
    Blue bands show 95\% confidence intervals of linear regression estimates.}
\end{center}
\end{figure*}

To determine whether we observe gender differences in \textit{longitudinal dynamics}, we compared the magnitude of the male and female valence curves at different times before and after the emotional expression using regression discontinuity analysis \cite{thistlethwaite:1960} (see Fig.~\ref{gender_difference}). 

In Fig. \ref{gender_difference}, we plot the difference of Valence values at time $t$ between male and female subjects, i.e. $V(t) = V_m(t) - V_f(t)$. Separate regression lines are calculated for $V(t)$ values before vs.~after $t_0$, including 95\% confidence intervals on the linear regression estimates. This allows us to determine whether male and female valence values evolve differently before or after the time of their emotional expression and whether the regression lines show statistically significant discontinuities.

As we observed earlier there is consistent and significant baseline difference between male and female valence (Male - Female $\in [-0.02,-0.10]$). Female subjects generally have higher valence baselines at all points before or after the emotional expression, for both positive and negative emotions.

At the time of an emotional statement, i.e. $t_0$, we observe discontinuities for both positive and negative emotions, indicating that the male and female valence diverge at the time of the expression.

This pattern is most pronounced for the positive valence time series. Prior to the emotional expression the gender difference increases (downward slope indicates that female valence increases more rapidly than male valence). This pattern is reversed immediately upon the emotional expression, indicating that female subjects experience a smaller reduction of positive emotions after the statement at time $t_0$ than male subjects do.

The negative valence time series indicates a constant baseline valence difference between male and female subjects \emph{before} the expression, indicating that the difference between male and female valence does not increase or decrease before the expression of a negative emotion. However, we again observe a discontinuity at the emotional expression at time $t_0$, although the CIs overlap slightly and we can not draw any firm conclusions about the statistical significance of the effect. Female subjects do seem to revert to baseline levels of negative valence emotions more slowly than male subjects after an emotional expression.

\section*{Discussion}

We report a methodological as well as a scientific innovation. First, we demonstrate the ability to use off-the-shelf natural language processing tools and large-scale social media data to detect how emotions evolve before and after they are explicitly expressed. Our results extend recent results~\cite{mason,ziemer2017} which suggest that sentiment analysis may reflect text sentiment, but not necessarily subject emotions. Assuming that our subjects correctly labeled their affect, our observations indicate that emotions may affect language hours before and after subjects express them explicitly.

Second, we empirically determine the parameters of how emotions change over time because of the evolving traces they leave in our subjects' language. Our results indicate that, depending on valence and gender, emotions last approximately 1.5 hours from onset to evanescence, with a decay half-life of about 11 and 5 minutes for positive and negative emotion respectively.

Third, our observations align with existing results in affect labeling studies that suggest the act of putting an emotion into words may lead to an attenuation of the emotion, without the subject explicitly recognizing or applying it as an emotional regulation strategy. We find that for a majority of our users emotional intensity decreases rapidly after their explicit expression in an ``I feel'' statement. This is the case for positive and negative emotions, but the effect seems to be strongest and most immediate for negative emotions. Finally, we find inconclusive, but suggestive indications that the evolution of positive and negative emotions may differentiate by gender, and that they differentiate most strongly at the time of the emotional statement.

Our method does have limitations. We rely entirely on a post-hoc, data-driven analysis. Although this reduces the possibility for some forms of observer bias, e.g.~subjects' emotions are changed by the act of reporting them, it is nevertheless possible that our sample is biased by particular social media characteristics such as recruitment and interface design. In addition, since social media platforms provide a public outlet, social drivers may need to be accounted for in future research, as well as a comparative analysis of different languages and possibly culture.  Our null-model is however carefully designed to account for many such biases by sampling from the respective positive and negative subject groups and taking into account diurnal and circadian rhythms. Our sentiment analysis tool responds only to English which may homogenize the linguistic and cultural variety of our sample.

Future investigations may require large-scale experiments that establish ground truth from \emph{in vivo} measurements, e.g. through surveys or facial musculature, which can be compared to computational indicators from large-scale social media data. These indicators can be expanded to study the dynamics of a variety of human emotions. Although valence is one of the primary components of most models of human emotions, our analysis could be expanded to include ``arousal'', ``dominance'', ``fear'', ``amusement'', ``calm'', and others which have recently been uncovered by an analysis of self-reports \cite{cowen:selfreport2017}. This will however require the design of specific natural language processing techniques geared towards the detection of such emotions in online language \cite{Bollen:twitter2011}.

Finally, a future analysis of our time series can be aimed at uncovering feedback processes in emotional dynamics to formulate a quantitative, mechanistic model of how language interacts with emotions and how individuals may self-regulate their emotional state, including the determination of individual and collective emotional resilience levels in terms of how quickly (groups of) individuals return to baseline levels after emotional perturbations. Our result lays the foundation for the development of mechanistic models of individual as well as population-level emotional dynamics. 

\section*{Methods}
\subsection*{Data and subjects} We randomly chose approximately 710,000 Twitter user IDs and issued requests to the Twitter's Application Program Interface (API) to retrieve all past tweets by the given user ID (up to a maximum of the 3,200 most recent tweets). The harvesting took place in 2012 hence the dates of our tweets ranges from 2006 (twitter starts) to 2012 (time of collection). For every user we obtained public profile information and for every tweet we retrieved a unique identifier, its content (140 characters), and the local UTC time at which it was posted. All timelines were stored in a MongoDB allowing fast and efficient programmatic access. Not all accounts queried were active or functional, so our final sample consisted of 665,081 subjects. To ensure our data pertained to individual users, we applied a series of filters described below.

\subsection*{Detecting self-reflective emotional statements}

We detect when subjects explicitly express or self-diagnose their emotional state at a particular point in time by stating either ``I feel ...'', ``I'm feeling ...'', or ``I am feeling''. We take the positive or negative valence of the adverb or adjective following the statement as an indication of the emotion’s valence, e.g. ``I feel \textbf{bad}'' means the subject is experiencing a negative valence emotion.

We use this explicit affirmation of a subjective emotional state as ground truth. For example, if a subject tweets: ``I feel bad'' at Friday 14:30EST we deem the subject to be truly in a negative emotional state at that specific point in time. Vice versa, if a subject tweets: ``I feel good'' Tuesday 09:30EST we deem the subject to be truly in a positive emotional state at that time.

The first step is to extract from our large set of timelines all tweets in which a subject clearly and explicitly affirms a positive or negative emotion by using the above listed ``I feel $\cdots$'' expressions. We define a restrictive``pattern'' of emotional affirmation to capture only the most explicit statements of having an high or low Valence emotion, namely

\begin{verbatim}
I feel|I'm feeling|I am feeling ($booster)? ($adjective|$adverb)
\end{verbatim}

The list of adverbs and adjectives that can follow these expressions are chosen to consist of the smallest possible yet most complete set of words that are frequently used in the English language to express positive or negative emotional states.

To achieve such a list, we start with the list of 13,916 Affective Norms of English Words~\cite{warriner2013norms}, each term rated by human subjects according to their Valence on a scale from 1 to 9. We identified 6,051 words having valence values higher than 6 and lower than 4 to remove words with neutral Valence values.  We subsequently identify all tweets containing the mentioned ``I feel'' patterns that also contained the sub-sample of non-neutral ANEW keywords. The ANEW term ``like'' was excluded from this analysis since it is most commonly used to express a figure of speech, e.g. ``I feel like a train hit me'' (53\% of all statements).

The keywords were rank-ordered by the frequency with which their ``I feel '' statements occurred in our data. Among the 50 most frequent keywords, two of the authors serving as independent annotators selected those adjectives best suited that expressed valence only, such as ``good'', ``great'', ``bad'' or ''horrible'', but not more specific emotions, e.g.~``fearful'' or ``confident'' or other psychological or physical states, e.g. ``fat'' and ``lazy''. Agreement between the annotators was 100\%. 

Of the selected adjectives, 4 indicated positive valence and 5 indicated negative valence. Since the positive valence terms contained ``happy'', ``unhappy'' was added to the list of negative valence adjectives for balance, even though the latter was not included in the list of the 50 most frequent adjectives (rank 282), bringing the total to 4 positive adjectives and 6 negative adjectives.

\begin{figure*}[h!]
\begin{center}
    \includegraphics[width=6in]{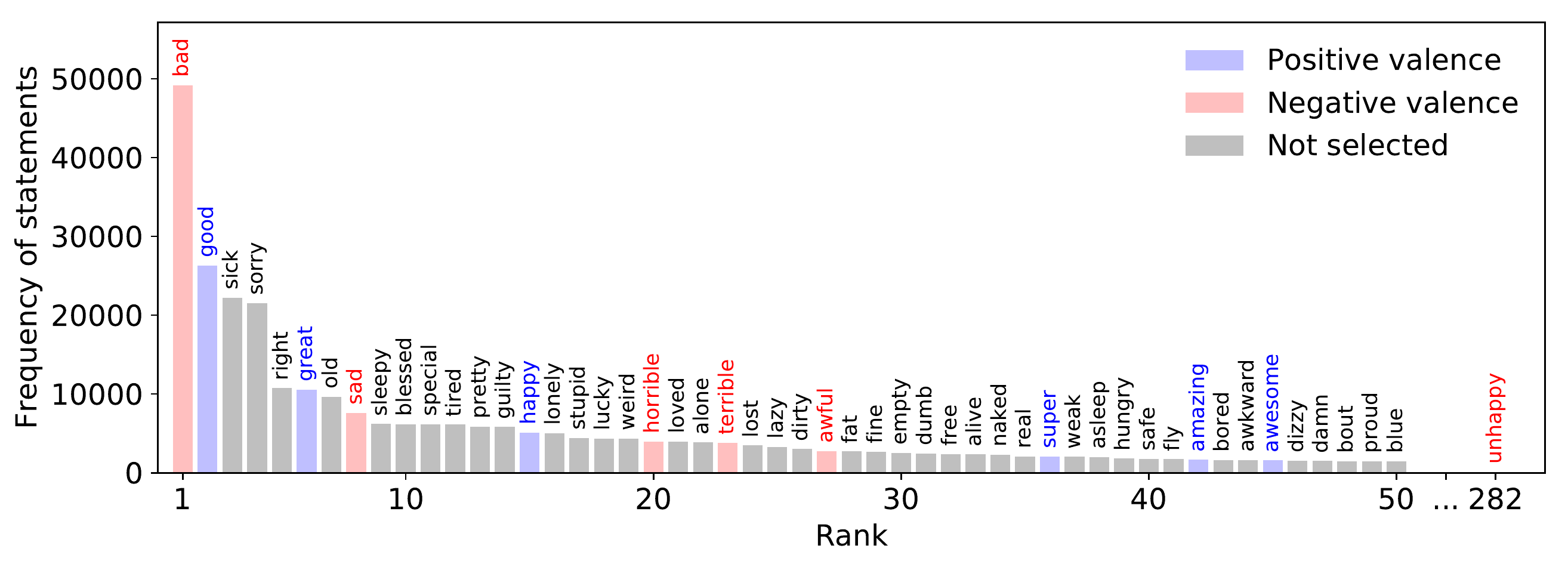}
    \caption{\label{word_ranking}``I feel '' adjectives rank-ordered according to their frequency of occurrence. Two annotators selected 10 adjectives that most unambiguously identified a low or high valence emotion, without connoting other more specific emotions (``sorry'', stupid'', ``weird''), physical states (``sick'', ``old''), figurative uses (``naked'', ``empty''), or having alternate uses as modifiers (``like'', ``pretty''). The positive and negative adjectives we identified are respectively shown in blue and red colors. Adjectives and other keywords that were not selected for the reasons above are show in gray.}
\end{center}
\end{figure*}

Our final list of valence adjectives for positive and negative affirmation consist of ``good'', ```happy'', `great'', ``awesome'' for positive emotions and ``bad'', ```unhappy'', `sad'', ``terrible'', ``horrible'', ``awful'' for negative emotions. Their corresponding frequency rank among emotional expression tweets is shown in Fig.~\ref{word_ranking}.

We allow each adjective to be preceded by a so-called ``booster'' word, for example ``I am feeling \emph{so} happy''. Our list of 46 booster words are taken from the VADER sentiment analysis tool~\cite{hutto2014vader}.

We label each emotional affirmation as ``negative'' or ``positive'' depending on the valence of the particular adjective or adverb it contains, i.e.~it indicates whether the subject at that point in time is experiencing a negative or positive emotion.

This procedure is focused on taking only a few of the most reliable, simple, and unambiguous emotional expressions in consideration, to reduce noise in our assessment of whether the subject experienced an emotion or not. The subsequent sentiment analysis (using VADER) of all other tweets (excluding the emotional statement) does account for negation, punctuation, and a variety of other linguistic features that are relevant to sentiment analysis. In total, we find 109,943 expressions of emotion, i.e.~42,627 positive ones and 67,316 negative ones.

To exclude further factors that could introduce noise in our analysis, we filter our timelines as follows.

\begin{itemize}
    \item We exclude Tweets marked as ``retweets'' since by definition they do not reflect the subject's own condition but that of those they have retweeted.
    \item To exclude users that are exceptionally prone to overshare emotions, we compute the proportion of emotional expression tweets for each user and then remove outliers who post more emotional expression than 95\% of all users.
    \item We remove timelines whose emotional expression tweets are posted on days that are unusual in terms of their number of emotional expression, i.e. days with less pattern tweets than 5\% or more emotional expression than 95\% of days in the year.
    \item We remove tweets from the timelines if they are not emotional expressions, but still contain ``i feel'', ``i am feeling'' or ``i'm feeling''. By doing this, we attempt to exclude false positives in our sentiment analysis.
    \item Given the prevalence of circadian rhythms \cite{Golder1878} and the fact that we are performing a longitudinal analysis that needs to take into account time of day, we remove timelines for users without timezone information.
    \item We remove cases in which the same subject posts more than one emotional expression in the same 48 hours period ($[t_0-24, t_0+24]$)d. By doing this, we increase the odds of each timeline in our data set representing a single, independent positive or negative emotional expression.
\end{itemize}

After these filters, 74,487 timelines remain of which most, i.e.~73,185 (98.25\%), are majority English. Each timeline contains 1 emotional expression tweet, and all tweets that conform to the above criteria that were posted by the same subject 6 hours before to 6 hours after the emotional expression. We refer to timelines that contain positive expression tweets as ``positive timelines'', the set of which is denoted as $D^+$. Vice versa, we refer to subject timelines that contain negative expression tweets as ``negative timelines'', the set of which is denoted as $D^-$. $D^+$ and $D^-$ contain 27,865 and 46,622 timelines respectively. By applying the gender classifier on subjects in these timelines, we achieve 9,382 male positive, 11,093 female positive, 13,118 male negative and 23,091 female negative timelines.

Subsequently, $D^+$ and $D^-$ can be analyzed to reveal temporal changes in the subjects' emotions as they approach or recede in time from the affirmation at $t_0$.

\subsection*{VADER sentiment analysis}

The VADER sentiment analysis tool was introduced by Hutto and Gilbert in 2014~\cite{hutto2014vader}. VADER is based on (1) an extended lexicon that contains over 7,500 lexical features and common online expressions and (2) a set of linguistic rules to deal with common grammatical features of online language. VADER was specifically tailored to rate social media content; its lexicon is augmented with emoticons, sentiment-related acronyms, and commonly used slang. The sentiment rating is conducted following five general rules to enhance its performance, e.g., recognition of punctuation such as exclamation points and capitalization increase rating intensity.

The F1 classification accuracy of VADER on Twitter data sets was found to be 0.96, which surprisingly exceeds that of individual human raters (0.84). A recent survey~\cite{Ribeiro2016} also suggested that VADER produces the highest accuracy sentiment rankings for a dedicated Twitter data set (F1\_pos=99.25, F1\_neg=98.33, Macro-F1=98.79, and  Coverage=94.61) among 22 tools, surpassing commonly used tools such as LIWC and SentiStrength. 

It is important to stress that the performance of VADER has been vetted in the context of text sentiment analysis, i.e.~whether a short text evinces a positive or negative sentiment, but not its ability to measure the emotions of the subject(s) that produce the text. We consider the latter a possible contribution of our research.

\subsection*{Gender classifier}
Men and women may differentiate with respect to how they experience and express emotions. However, Twitter does not provide reliable gender information about its users, other than what they may sporadically self-report.

We therefore split our sample of subjects into a male and female group by a gender classifier as follows. Following~\cite{li2014weakly}, we extract gender labels from Google Plus profiles where the subjects have specified their genders and also include a link to their Twitter account. Then, we manually verify the genders of 1,744 subjects, including 658 female and 1,086 male subjects respectively. We collect their Twitter timelines using the free Twitter API to build our training data.

Each timeline is represented with a 200 dimensional vector extracted from the GloVe~\cite{pennington2014glove} data set which consists of low-dimensional vector word representations pre-trained from 2 Billion tweets as follows. First, each timeline is treated as one document and words with document frequency of less than 30 are removed. We retain the top 100 words features by using an ANOVA F-value feature selection method. At last, we use the TF-IDF method to assign weights to each word and build the vector of each timeline by weighted summing of all GloVe vectors of the selected features in the timeline.

Based on the vector representation of each timeline, we train a binary gender classifier using a calibrated Random Forest method~\cite{Breiman2001,niculescu2005predicting} with 160 estimators, producing the probability of being female for the target subject. The classifier performs fairly well (accuracy=86.4\%, AUC score=0.916), surpassing commonly used SVM and Na\"ive Bayes methods. When applying the classifier to our subjects, to make the results more reliable, we use 0.3 and 0.7 as thresholds for male and female respectively, i.e. if the probability produced by the classifier is smaller than 0.3, we label the corresponding subject as male and if it's larger than 0.7, we label the subject as female.

With respect to accuracy, note that incorrect gender predictions will lead to mixed samples and reduce the magnitude of any observed male-female differences. Hence the gender effects we report are likely underestimations of the true effect, as our "male" group may contain females and vice versa.

\subsection*{Null-model}

We build a null-model to determine the statistical significance of our sentiment values for each window. We follow the procedure described below to construct a null-model:

\begin{itemize}
    \item Subject dispositions: We randomly sample tweets from positive and negative subject timelines separately to construct a positive and negative null-model to account for subject dispositions (subjects expressing negative emotions might be overall more negative than subjects expressing positive emotions).
    \item Diurnal cycles: We randomly sample from the 24 hours surrounding $t_0$ to account for bias associated with diurnal cycles and its effect on the timing of emotional expression.
    \item Circadian and weekly cycles: Our random sample is selected to match the distribution of time of day and week day (Monday, Tuesday, etc) distribution of the observed sample, since some emotional expressions might be more common or biased at certain times of day or certain weekdays (e.g. Monday morning vs. Friday night).
    \item Sample size: For each time window, we sample the \emph{same} number of tweets as we observed for that window. Therefore, as shown in Fig. \ref{timeseries_KS10}, CI bands narrow towards $t_0$, since the volume of tweets increases and uncertainty of our mean Valence estimates thus decreases.
    \item Computing mean valence and confidence intervals: We sample the \emph{same number} of tweets as observed in each window and calculate mean valence for the window. We repeat this sampling procedure 10,000 times (with replacement) and compute 5th, 50th, and 95th percentiles for the resulting distribution of mean Valence values to obtain confidence intervals for our null-model.
\end{itemize}

Note that our procedure results in a null-model that is \emph{more strict} than an entirely random sample of all tweets and all subjects, since it requires that any Valence change observed in a given time window diverges significantly from one that is expected by chance for the \emph{same cohort} of positive and negative subjects, the \emph{same number} of tweets observed, \emph{same time} of day, \emph{same week day}, and within the \emph{same circadian and diurnal cycle} as the tweets we observe in our timelines.

\subsection*{CUSUM test}

\begin{figure*}[h!]
\begin{center}
    \includegraphics[width=6in]{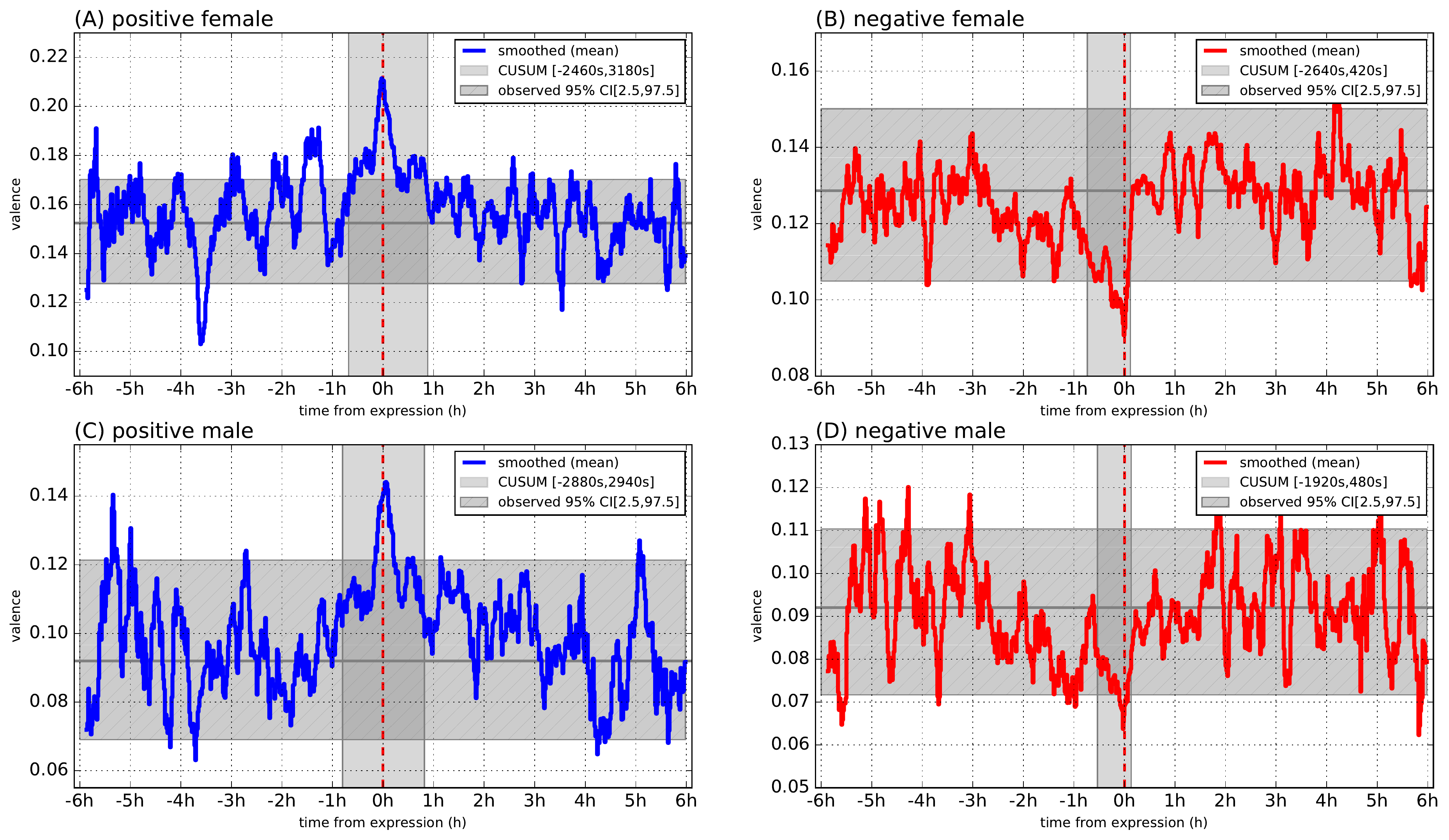}
    \caption{\label{timeseries_simple_femmal} Gender differentiated time series of mean valence values in adjacent 1 minute windows before and after an explicit expression of having a positive or negative emotion at time $t_0$. Time series are smoothed by a 10 minute rolling average. The blue and red bands shows a 95\% confidence intervals of the distribution of valence values determined from the first 3 hours of the time series. A CUSUM test reveals where statistically significant changes in the time series occur (vertical gray bands).}
\end{center}
\end{figure*}

We apply the CUSUM anomaly detection method on the $t_0 \pm 6$hours smoothed mean time series. First, we calculate the CUSUM chart which contains the upper control limit series by $S^+_i=\max{[0, S^+_{i-1}+x_i-(T+K)]}$ and the lower control limit series by $S^-_i=\min{[0, S^-_{i-1}+x_i-(T-K)]}$. $x_i$ represents the series value at $i$. $T$ and $K$ are the mean and standard deviation of the corresponding null-model series, reflecting the expected mean and standard deviation of the median score series. Then points that satisfy $S^+_i>H$ and $S^-<-H$ ($H=0.01$) are chosen as the upper violations and lower violations respectively. Anomalies are detected by picking the increasing sub-sequences of the upper violations and decreasing sub-sequences of the lower violations. Sequences that are longer than $\lambda=40$min are determined to be our emotional periods, meaning that in these periods subjects' emotions are significantly higher or lower than chance levels. In Fig~\ref{timeseries_simple_femmal}, we present the CUSUM analysis for gender differentiated time series. 

\subsection*{Data and code availability statements}

\textbf{Data}: The Twitter content data that support the findings of this study are publicly available from Twitter, but can not be distributed by the authors. The authors will however provide the Twitter identification codes of all tweets used in this analysis to allow for retrieval of their content from the Twitter API. All other data are available from the authors upon reasonable request.

\section*{Acknowledgments}
R. F. thanks the support from NSFC (Grant No. 71501005). Johan Bollen thanks the National Science Foundation (SMA-SBE: 1636636), Wageningen University (The Netherlands), and the ISI Foundation (Turin, Italy) for their support. Special thanks to Filippo Radicchi and Camila Scheffer for their insightful comments, and Rion B. Carreia who kindly set up our data bases. The authors also thank Bruno Goncalves for his efforts. We would also like to thank our reviewers whose constructive feedback helped us to improve previous versions of this paper.

\bibliographystyle{plain}
\bibliography{realmood}

\end{document}